\documentclass[preprint,12pt]{elsarticle}

\usepackage{amsmath,amssymb,array,calc,psfrag,empheq}
\usepackage{graphicx}
\usepackage{caption}
\usepackage{subcaption}

\title{Analysis of intra-day fluctuations in the Mexican financial market index}

\author{-L\'ester Alfonso$^a$,
	Danahe E. Garcia-Ramirez$^b$, 
	Ricardo Mansilla$^c$, C\'esar A. Terrero-Escalante$^d$\corref{cor1}}
\cortext[cor1]{Author's names are arranged alphabetically. Corresponding author: cterrero@ucol.mx}
\address{$^a$ Universidad Aut\'onoma de la Ciudad de M\'exico,\\ C.P. 09790, M\'exico D.F.,  M\'exico\\
$^b$ Departamento de Astronom\'ia,\\ 
DCNE-CGT, Universidad de Guanajuato,\\ C.P. 36023, Guanajuato,~M\'exico.\\
$^c$ Centro de Investigaciones Interdisciplinarias en Ciencias y Humanidades,\\ 
Universidad Nacional Aut\'onoma de M\'exico,\\ 
Ciudad Universitaria, C.P. 04510, M\'exico~D.F.,~M\'exico.\\
$^d$Facultad de Ciencias, Universidad de Colima,\\
Bernal D\'\i az del Castillo 340, Col. Villas San Sebasti\'an,\\ 
C.P. 28045, Colima, Colima, M\'exico.\\
}

\journal{Physica A}
\date{\today}

\begin{document}

\begin{frontmatter}

\begin{abstract}
In this paper, a statistical analysis of high frequency fluctuations of the IPC, the Mexican Stock Market Index, is presented. 
A sample of tick--to--tick data covering the period from January 1999 to December 2002 was analyzed, as well as several other sets obtained using temporal aggregation.
Our results indicates that the highest frequency is not  useful to understand the Mexican market because almost two thirds of the information corresponds to inactivity. 
For the frequency where fluctuations start to be relevant,
the IPC data does not follows any $\alpha$-stable distribution, including the Gaussian,
perhaps because of the presence of autocorrelations.
For a long range of lower-frequencies,
but still in the intra-day regime,
fluctuations can be described as a truncated L\'evy flight,
while for frequencies above two-days,
a Gaussian distribution yields the best fit.
Thought these results are consistent with other previously reported for several markets,
there are significant differences in the details of the corresponding descriptions.  
\end{abstract}


\end{frontmatter}

\section{Introduction}
\label{sec:Intro}

Is a fundamental assumption of the economic theory of markets that the financial markets are efficient in the sense of the community of economic agents being able to discount all the possibilities of arbitrage, incorporating with it all the relevant information for the prices formation,
so that it is impossible to design an (always) winning strategy for investment\cite{fama, mantegna-stanley-book}.
\footnote{For more details on the background and definitions of the concepts we use in this paper see book \cite{mantegna-stanley-book}.} 
Mathematically, this 
is described as a process, where
a future increase or decrease of the current price
is the result of a random event.
Underlying such a random walk is a binomial distribution, which after a very large number of steps (price changes), converges to a normal, Gaussian, distribution.
In turn, normal distributions are common in the description of systems in equilibrium,
a stable state, small fluctuations around which decay exponentially with time. 
However, it seems that the financial market cannot actually be modeled as this kind of system; 
data for most financial indexes around the world are statistically described by probability distributions that exhibit a large skewness or kurtosis, relative to that of a normal distribution. 
It may means that strong perturbations are not, 
de facto, 
necessary for the market entering a critical state;
recurrent significant deviations of economic variables from their average values
could be just the cumulative and unavoidable result of many small-scale processes and interactions occurring within the market system,
continuously subjected to a net external action like,
for instance,
publicly available announcements of annual earnings, stock splits, companies profits forecasts, new securities or,
even more,
the punctual action of agents having preferential access to restricted or confidential information. 
 
In this regard, for years now particularly good fits to financial data have been obtained using L\'evy-stable distributions \cite{levy}.
Nevertheless, it is obvious that, strictly speaking, the market can neither be such a random process. 
First, 
it is unreasonable to expect data from any real
(as opposed to hypothetical) 
process to have an infinite variance.
Secondly,
it is reasonable to expect data from financial transactions
at a given time to have some memory of the previous transactions,
i.e.,
the elements of the corresponding time series should not be really independent each from the others.
This is easily noted, for instance, during steeped variations
of the stock returns due to panic or euphoria.

Commonly there are two causes of autocorrelation in this kind of data, 
irregular sampling and causality determined by market microstructure.
In agreement with the ‘efficient market
hypothesis’,
it is intuitive not to expect persistent serial dependence in price changes, 
otherwise it would be used to influence the market.
Consistently with this, autocorrelation is often neutralized after homogenization of the corresponding series by time averaging of subsamples.
This typically results in sets with information correlated with those of daily or lower frequency data.
It is for these last sets that fitting to stable distributions have already yielded remarkably good results
\cite{mandelbrot,mantegna-stanley,gopikrishnan,cont, alfonso}.
More precisely, it has been verified that the actual density distribution behind the data could be
a so-called L\'evy-truncated \cite{mantegna-stanley-PRL,koponen} 
which, after sequential convolution 
corresponding to the sum of the values over increasing time intervals, 
also converges to a Gaussian distribution. 
This convergence is ultra-slow,
therefore the truncation still allows for a relatively high probability of extreme fluctuations
which,
as it was already noted, 
seems to be a distinctive feature of most financial markets.

Despite the success in describing markets activity as a L\'evy (truncated) flight,
it is widely acknowledged that the description
will not be complete unless high-frequency data is also brought into the study.
On one hand,
most market models are based on a variety of
hypothesis regarding the long-memory features of volatility
which are difficult to extract from daily or lower-frequency data,
but can be observed in intraday data.
For instance,
as it was already mentioned,
crisis in finances may not necessarily be linked
only to strong perturbations over several days,
but also to the cumulative effect of intraday weaker perturbations. 
On the other hand, 
the quality of risk analysis of investment depends on the accuracy of measurement of ex post volatility and of forecast evaluation. 
There is evidence of the improvement in both directions
thanks to the
availability of high-frequency data from liquid financial markets
such as the foreign-exchange, bond or equity-index markets afford \cite{andersen}.
Last, but not least,
the analysis of the impact of sample size on the probability estimation of extreme events yields that a large volume of data is needed in order to determine the actual stable distribution corresponding to a given asymptotic scaling\cite{weron,alfonso}.
This is the kind of volume that commonly characterizes the sets of high frequency data.

It must be pointed out that 
market microstructure and short term interactions become relevant
while analyzing intra-day fluctuations
and, since they depend on local socio-economic factors,
the stylized facts of the market as a complex system can be difficult to extract.
Therefore, it is very important to study different realizations of high frequency financial data
in order to discriminate local and universal properties of the market dynamics.
It is with this intention that in this paper we report a first step into analyzing
the tick-to-tick data of the Indice de Precios y Cotizaciones (IPC),
the main benchmark stock index in Mexico.
It has been previously shown in Ref.\cite{alfonso} that
it can be dismissed that a normal distribution relies under the data of daily closure values of the IPC. 
On the other hand, the null hypothesis that it comes from $\alpha$-stable L\'evy distribution cannot be rejected at the 5\% significance level.
This implies that the daily data can safely being considered as independent and identically distributed,
characterized by an infinite variance.
Our aim here is to study how this conclusion changes when tick-to-tick data is analyzed.

The paper is organized as follows. 
In the next section we provide a brief review of L\'evy distributions, in particular of the stable and L\'evy truncated distributions which we use for fitting the IPC fluctuations.
The description of this data, as well as the sets derived from them and used for our analysis is presented in section \ref{sec:data}.
Next in sections \ref{sec:stats} and \ref{sec:autocorr} we proceed to present the analysis of the probability distribution and serial dependence of the actual fluctuations of the IPC data.
We devote the last section to discuss our results and state the main conclusions. 

\section{L\'evy distributions}
\label{sec:levy}

As it was mentioned in the introduction,
L\'evy distributions are among the more frequently used
to fit data from complex processes.
Particularly, it has been found to be an
excellent fit to the distribution of stock returns
as well as of other financial time series.
In this section we will briefly review the definition and
properties of the stable and truncated
L\'evy distributions. 

\subsection{Stable distributions}
\label{ssec:stable}

While studying the behavior of sums of independent random variables Paul L\'evy \cite{levy} introduced
a skew distribution specified by scale $\gamma$, exponent $\alpha$, skewness parameter $\beta$ and a location parameter $\mu$. 
Since the analytical form of the Levy stable distribution is known only for a few cases, 
they are generally specified by their characteristic function. 
The most popular parameterization is defined by Samorodnitsky and Taqqu \cite{taqqu} with the characteristic function:
\begin{equation}
  \phi(t)=\begin{cases}
    \exp{\left(-\gamma |t| \left[1+i\beta\frac2{\pi }\rm{sign}(t) \ln(|t|) + i\mu t\right]
\right)}, & \text{if $\alpha=1$}.\\
    \exp{\left(-\gamma^\alpha |t|^\alpha \left[1-i\beta\tan\left(\frac{\pi \alpha}2\right) {\rm sign}(t) + i\mu t\right]
\right)}, & \text{otherwise}.
  \end{cases}
\end{equation}
where $sign(t)$ stands for the sign of $t$. 
Then, the probability density function is calculated from it with the inverse Fourier transform in the form: 
\begin{equation}
f(x;\alpha,\beta,\gamma,\mu)=\frac1{2\pi}\int^{+\infty}_{-\infty}\phi(t){\rm e}^{-itx} dt  \, .
\end{equation}	

L\'evy distributions are characterized by the property of being stable under convolution, 
i.e, the sum of two independent and identically L\'evy-distributed random variables, is also L\'evy distributed with the same stability index $\alpha$. 	
The stability parameter $\alpha$ lies in the interval $(0, 2]$. 
Small $\alpha$ represents a sharp peak but heavy tails which asymptotically decay as power laws with exponent $-(\alpha+1)$. For the normal distribution $\alpha=2$. 
For symmetric distribution (like the normal distribution), the skewness parameter $\beta=0$. The skewness parameter must lie in the range $[-1, 1]$. 
When $\beta=+1,-1$, one tail vanishes completely. 
The parameter $\gamma$ lies in the interval $(0, \infty)$, 
while the location parameter $\mu$ is in $(-\infty,+\infty)$.

The asymptotic behavior of the L\'evy distributions is described by the expression
\begin{equation}
f(x;\alpha)\approx |x|^{-1-\alpha} \, .
\end{equation}	                                                                                                                   
Hence, the variance of the Levy stable distributions is infinite for all $\alpha<2$. 

\subsection{Truncated L\'evy distributions}
\label{ssec:truncated}

As mentioned in the previous subsection, $\alpha$-stable Levy distributions have infinite variance, hence, they have power-law tails that decay too slowly. 
Therefore, the fit of empirical data by L\'evy stable distributions will usually overestimate the probability of extreme events. 
In particular, real prices fluctuations 
have finite variance,
so their distribution decays slower than a Gaussian, but faster than a Levy-stable distribution, 
with the tails better described by an exponential law
than by a power law. 
The truncated L\'evy flight (TLF) was proposed by Mantegna and Stanley \cite{mantegna-stanley-PRL} to overcome this problem, 
and can be defined as a stochastic process with finite variance and scaling relations in a large, but finite interval. 
They defined the truncated L\'evy flight distribution as:
\begin{equation}
P(x) =
\left\{
\begin{array}{lll}
	0  & \mbox{if } x > l \\
	c P_L(x) & \mbox{if }  -l \leq x \leq l \\
	0  & \mbox{if } x < -l \, ,
\end{array}
\right. 
\label{eq:truncated}
\end{equation}
where $P_L(x)$ is a symmetric L\'evy distribution. As the TLF has a finite variance, with sequential convolution it will converge to a Gaussian process, but the convergence is very slowly, as was demonstrated by Mantegna and Stanley \cite{mantegna-stanley-PRL}. 
However, the cutoff in the tail given by (\ref{eq:truncated}) is abrupt. 
This problem was solved by Koponen \cite{koponen}, who introduced an infinitely divisible TLF with an exponential cutoff with the characteristic function:
\begin{equation}
\log\phi(t) =- \frac{c^\alpha}{\cos(\frac{\pi\alpha}{2})}
\left[
\left(t^2+\lambda^2\right)^\frac{\alpha}{2}
\cos\left(\alpha \arctan\frac{|t|}{\lambda} 
- \lambda^\alpha\right)
\right] \, ,
\label{eq:koponen}
\end{equation}
where $c$ is the scaling factor,  $\alpha$ is the stability index, and $\lambda$ is the cutoff parameter. 
The L\'evy $\alpha$-stable law is restored by setting to zero the cutoff parameter. 
For small values of $x$, the truncated Levy density described by the characteristic function (\ref{eq:koponen}), behaves like a L\'evy-stable law of index $\alpha$ 
\cite{cont-potters-bouchaud}.
It was used by Matacz \cite{matacz} to describe the behavior of the Australian All Ordinaries Index.

\section{Data sets from IPC values}
\label{sec:data}

\begin{figure}[!ht]
	\centering
	\includegraphics[width=.7\linewidth]{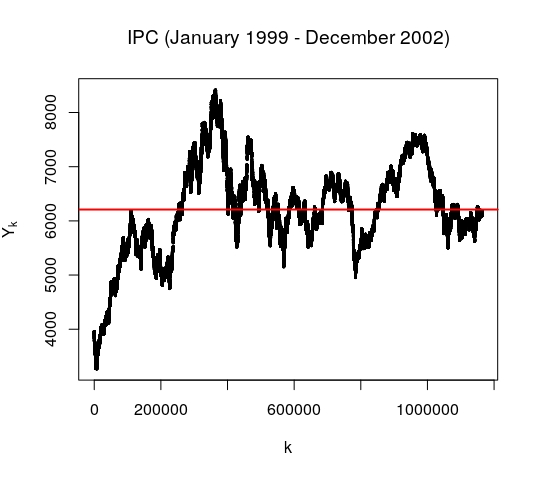}
	\caption{IPC actual fluctuations. The horizontal red line corresponds to the mean value over the period.}
	\label{fig:yk}
\end{figure}

For our analysis we used the IPC value over the period January 1999-December 2002 which comprises $4321427$ transactions. 
Taking into account that the Mexican trading day is of six and a half hours, this give us a mean time between transactions of $5.2$ seconds.
With the aim of analyzing the statistics of nontrivial index fluctuations,
the repeated values
(the intervals where no change in the returns was recorded)
were removed and
the set was reduced to $N=1164256$ elements
with a mean value over the period of $6209.392$,
a variance of $744334.3$
and an excess of kurtosis of $1537.442$.
These ticks, $Y_k$, are now irregularly sampled,
with a mean time between fluctuations of 19.3 seconds.
This set is plotted in figure \ref{fig:yk}.

The actual fluctuations are then taken as the difference of the natural logarithm of the $Y_k$,
\begin{equation}
S_k=\ln Y_{k+1} - \ln Y_k\, , \quad {\rm for}\, k=1,2,\dots, N \, .
\end{equation}

For this study we also used sets corresponding to the convolution of the density distributions of high-frequency data $\{S_k\}$, 
i.e.,
sets obtained after summing for different values of $N_{conv}$,
\begin{equation}
S_j^{N_{conv}}=\sum_{k=1+(j-1)\times N_{conv}}^{j\times N_{conv}} S_k\, , \quad {\rm for}\, j=1,2,\dots, \bar{N} \, ,
\end{equation} 
where $\bar{N}$ is the multiple of $N_{conv}$ closer to $N$.

For completeness we analyzed also a set of closing values of the IPC for the same period downloaded from the Yahoo Finance website.

\section{Probability distribution of IPC fluctuations}
\label{sec:stats}

\begin{figure}[!ht]
\centering
\includegraphics[width=.9\linewidth]{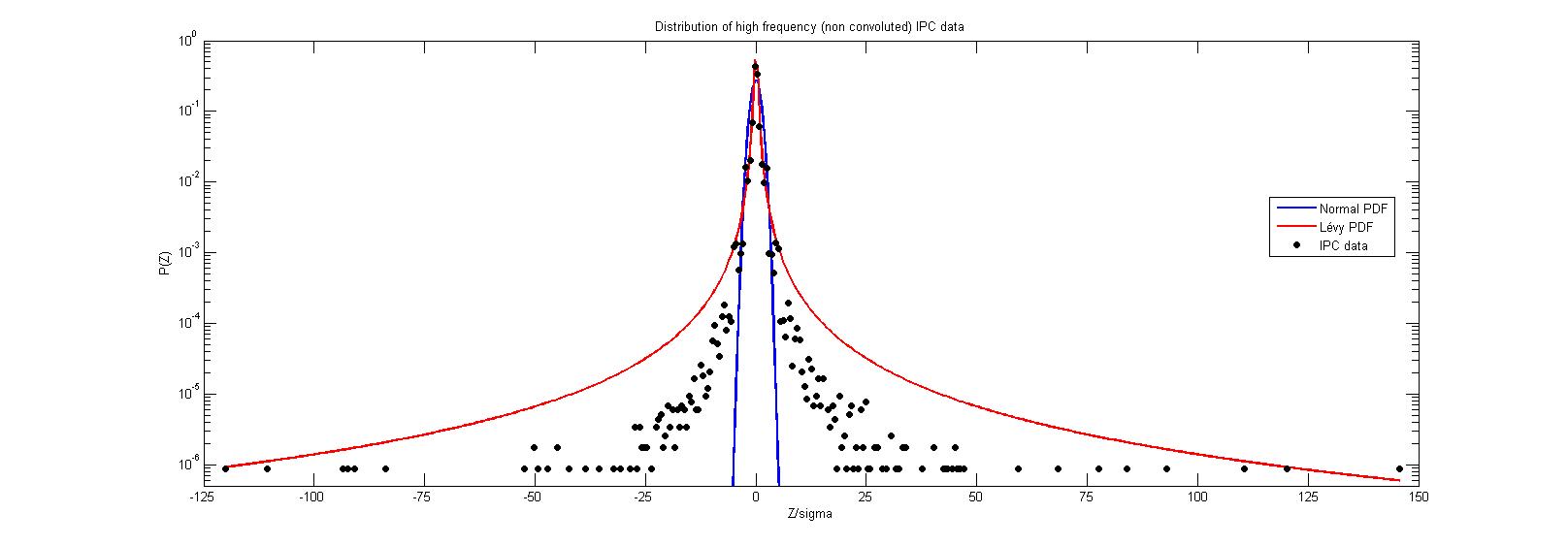}
\caption{Distributions for $N_{conv}=0$. In the horizontal axis $Z$ stands for the fluctuations data or the values of the corresponding fits.}
\label{fig:distconv0}
\end{figure}

We started by analyzing the non-convoluted data.
The corresponding distribution is presented in figure \ref{fig:distconv0},
where a logarithmic scale is used for the vertical axis
and the horizontal axis has been rescaled dividing by the standard deviation.
The plots of best fits to a Gaussian
(narrow blue curve below the data points)
and a L\'evy 
(wide red curve above the data points)
distributions are also shown.\footnote{For the analysis of stable distributions, we used the library developed by J. Nolan 
\cite{nolan}.}
It can be observed that the probability distribution function for this data 
does not correspond to a normal distribution,
but it neither does to a L\'evy one.
This is confirmed by the results of the Kolmogorov-Smirnov test shown
in table \ref{tab:table1}
for the best fit of data to an $\alpha$-stable distribution.
\begin{table}
\caption{Results of the Kolmogorov-Smirnov Goodness of Fit Test (K-S test).}
\centering
\begin{tabular}{|l|c|c|c|c||c||c||c|}
\hline
$N_{conv}$ & $\alpha$ & $\beta$ & $\gamma$ & $\delta$ & K-S Statistics & p-value 
& \multicolumn{1}{|p{2cm}|}{\centering Reject $H_0$? $p=0.05$\\ } \\
\hline
0  & $1.2565$  & $-0.0024$ &  $0.3796$  & $0.0014$ & $0.0179$ & $0.0$ & Yes \\
\hline
10  & $1.5788$  & $0.0056$ &  $2.3138$  & $-0.0051$ & $0.0097$ & $0.0$ & Yes \\
\hline
20  & $1.6107$  & $0.0089$ &  $3.8497$  & $-0.0114$ & $0.0076$ & $0.0026$ & Yes \\
\hline
30  & $1.6296$  & $0.0072$ &  $5.2420$  & $-0.0075$ & $0.0081$ & $0.0115$ & Yes \\
\hline
40  & $1.6539$  & $0.0140$ &  $6.6326$  & $-0.03553$ & $0.0099$ & $0.0064$ & Yes\\ 
\hline
50  & $1.6523$  & $0.2332$ &  $7.8216$  & $-0.0497$ & $0.0108$ & $0.0084$ & Yes\\ 
\hline
60  & $1.6594$  & $0.0145$ &  $8.9391$  & $-0.0492$ & $0.0100$ & $0.0404$ & Yes\\ 
\hline
70  & $1.6647$  & $-0.0053$ &  $10.0793$  & $0.0359$ & $0.0078$ & $0.2602$ & No \\ 
\hline
80  & $1.6682$  & $0.0010$ &  $11.1893$  & $0.0339$ & $0.0089$ & $0.2022$ & No \\ 
\hline
90  & $1.6688$  & $0.0141$ &  $12.1705$  & $0.0103$ & $0.0102$ & $0.1365$ & No \\ 
\hline
100  & $1.6845$  & $0.0029$ &  $13.2942$  & $0.0329$ & $0.0099$ & $0.2053$ & No \\ 
\hline
110  & $1.6916$  & $0.0080$ &  $14.2635$  & $-0.0547$ & $0.0123$ & $0.0819$ & No \\ 
\hline
120  & $1.6887$  & $0.0049$ &  $15.2096$  & $-0.0066$ & $0.0114$ & $0.1607$ & No \\ 
\hline
130  & $1.6826$  & $-0.0291$ &  $16.0270$  & $0.2303$ & $0.0131$ & $0.0903$ & No \\ 
\hline
140  & $1.6874$  & $-0.0211$ &  $16.9094$  & $0.2623$ & $0.0115$ & $0.2163$ & No \\ 
\hline
150  & $1.6909$  & $-0.0027$ &  $17.8762$  & $0.1131$ & $0.0107$ & $0.3344$ & No \\ 
\hline
1200  & $1.8693$  & $-0.3670$ &  $71.8964$  & $6.8822$ & $0.0150$ & $0.5328$ & No\\ 
\hline
2500  & $1.9065$  & $-0.6018$ &  $108.4828$  & $12.0903$ & $0.0301$ & $0.7860$ & No\\ 
\hline
2700  & $2.0000$  & $0.7973$ &  $122.6070$  & $5.0104$ & $0.0309$ & $0.7985$ & No\\ 
\hline
\end{tabular}
\label{tab:table1}
\end{table}

\begin{figure}[!ht]
\centering
\includegraphics[width=.9\linewidth]{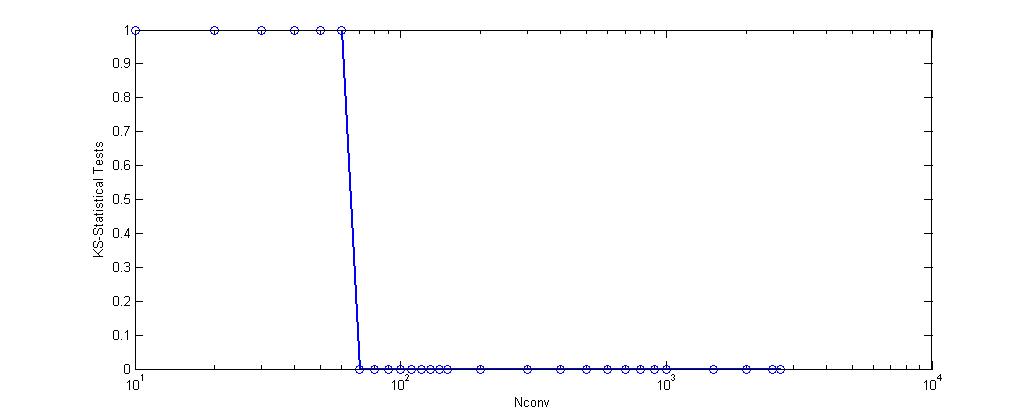}
\caption{Results of the Kolmogorov-Smirnov Goodness of Fit Test versus $N_{conv}$.}
\label{fig:KStest}
\end{figure}

\begin{figure}[!ht]
\centering
\includegraphics[width=.9\linewidth]{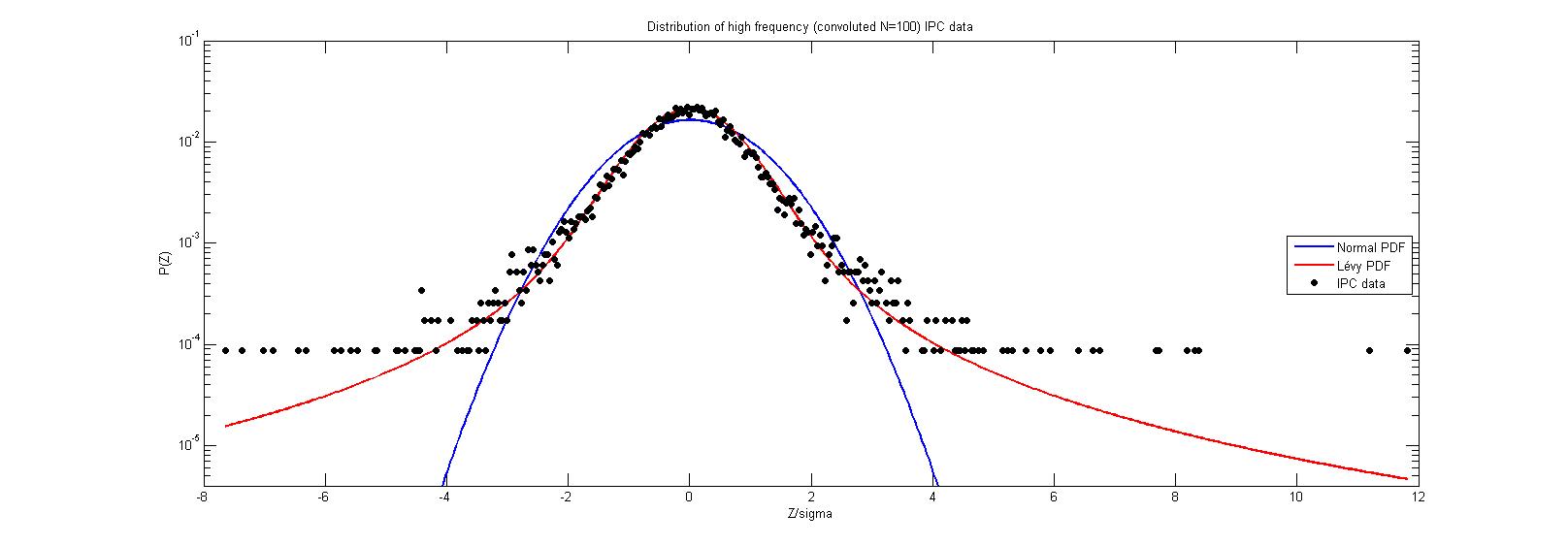}
\caption{Distributions for $N_{conv}=100$. In the horizontal axis $Z$ stands for the fluctuations data or the values of the corresponding fits.}
\label{fig:distconv100}
\end{figure}

From this table we can also observe that as $N_{conv}$ reaches a value
around $70$, the test cannot longer reject the hypothesis of the probability 
density function for this data being an $\alpha$-stable distribution.
As it is shown in fig.\ref{fig:KStest},
L\'evy scaling holds over a long range of values of $N_{conv}$.
For instance, 
in fig.\ref{fig:distconv100} is presented the data 
and the best fits for a normal and an $\alpha$-stable distribution
for $N_{conv}=100$.

\begin{figure}[!ht]
	\centering
	\includegraphics[width=.9\linewidth]{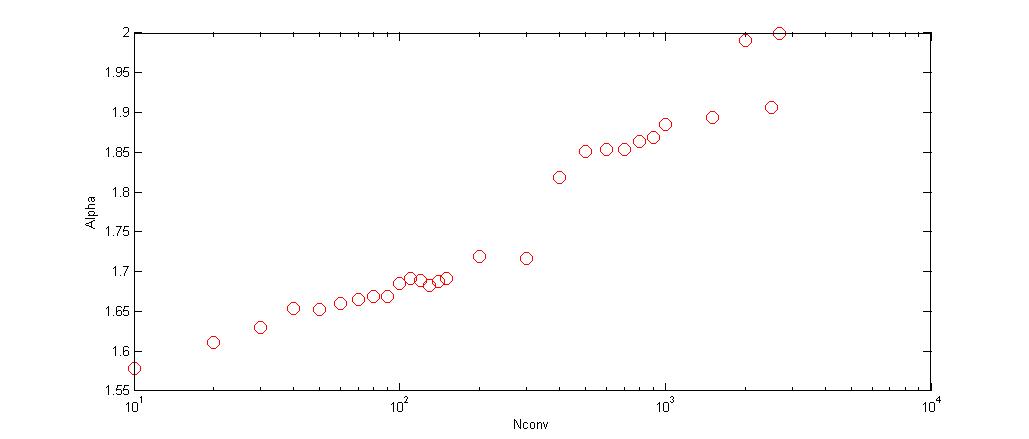}
	\caption{The stable coefficient $\alpha$ as function of $N_{conv}$.}
	\label{fig:alphavsNconv}
\end{figure}

\begin{figure}[!ht]
\centering
\includegraphics[width=.9\linewidth]{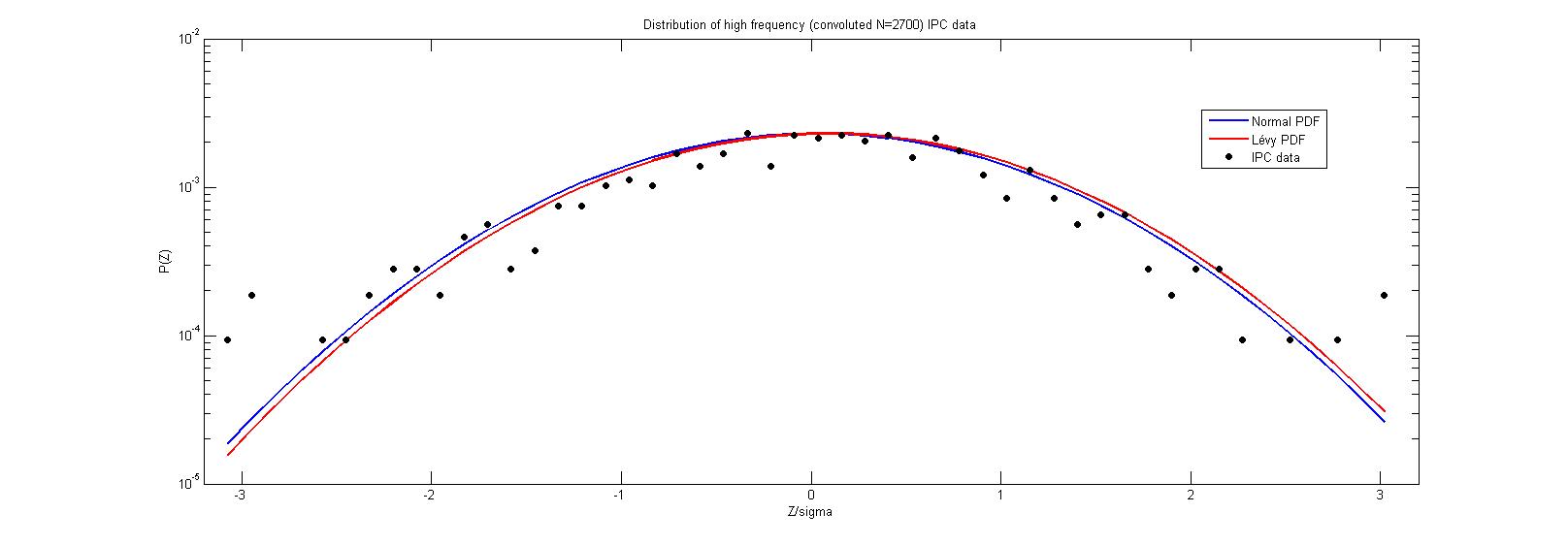}
\caption{Distributions for $N_{conv}=2700$. In the horizontal axis $Z$ stands for the fluctuations data or the values of the corresponding fits.}
\label{fig:distconv2700}
\end{figure}

The values of $\alpha$ keep steadily increasing as $N_{conv}$ is also increased.
As a matter of fact,
as it can be seen in table \ref{tab:table1}
and it is represented in Fig.\ref{fig:alphavsNconv},
the value of the stable coefficient slowly converges to
$2$,
while the convolution involves larger blocks of data.
That is, for instance, the value of $\alpha$ for $N_{conv}=2700$.
For this case the corresponding statistics
are presented in the last row of table \ref{tab:table1}
and the distributions are plotted in Fig.\ref{fig:distconv2700}.
As we can see,
for $N_{conv}=2700$
a normal distribution
and the corresponding L\'evy distribution
(with $\alpha=2$)
give both a very good fit to the convoluted data.
This is a strong evidence that,
indeed,
the variance of the data is finite.

\subsection{Convergence to $\alpha=2$}
\label{ssec:convergence}

\begin{figure}[!ht]
\centering
\includegraphics[width=.7\linewidth]{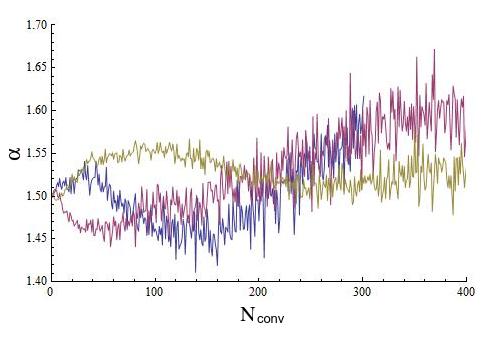}
\caption{The stable coefficient $\alpha$ as function of $N_{conv}$, without truncation.}
\label{fig:sintruncar}
\end{figure}

Note in figure \ref{fig:alphavsNconv} that $\alpha\approx 2$ 
for $N_{conv}$ around $2000$ too.
Also in table \ref{tab:table1} 
it can be seen that the convergence to $\alpha=2$
is not just slow,
as has been noted previously \cite{mantegna-stanley-PRL},
but it is also non-uniform.
We believe that this is an effect of the finite number of elements in the sample.
To verify that, 
we simulated data by truncating sets generated using the
stable library by Nolan \cite{nolan} 
and following expression (\ref{eq:truncated}).
For the three sets we used the same parameters,
but the length of the series are one, two and five millions elements respectively.
In fig.\ref{fig:sintruncar} is plotted 
how the stable coefficient $\alpha$ for each original set (without truncation)
evolves with convolution.
It is noticeable the high quality of the simulated data,
since in each case $\alpha$ converges to a value significantly different
from $2$ and,
the larger the size of the set,
 the closer this value gets to the $\alpha$ used for generating the set.

\begin{figure}[!ht]
\centering
\includegraphics[width=.7\linewidth]{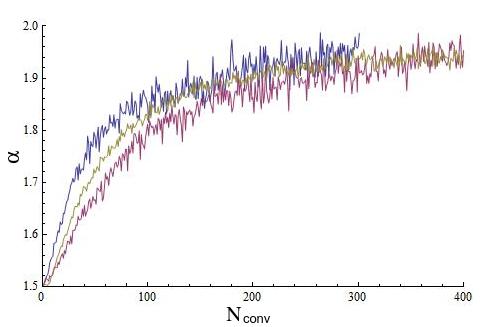}
\caption{The stable coefficient $\alpha$ as function of $N_{conv}$ for $n_{std}=50$.}
\label{fig:50std}
\end{figure}

\begin{figure}[!ht]
\centering\ref{fig:alphavsNconv}
\includegraphics[width=.7\linewidth]{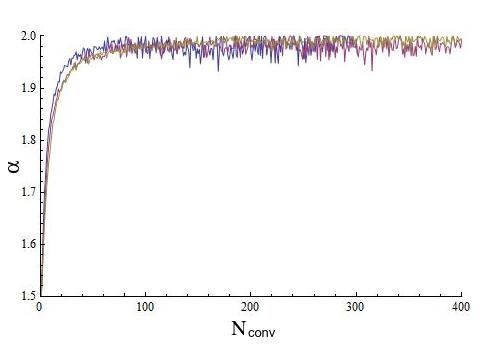}
\caption{The stable coefficient $\alpha$ as function of $N_{conv}$ for $n_{std}=10$.}
\label{fig:10std}
\end{figure}

\begin{figure}[!ht]
\centering
\includegraphics[width=.7\linewidth]{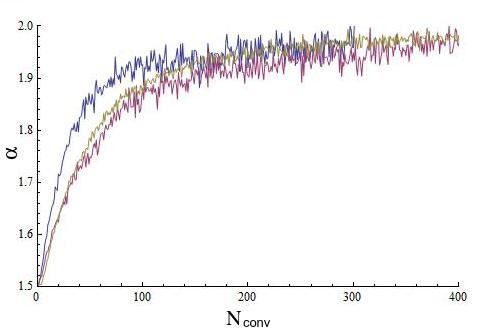}
\caption{The stable coefficient $\alpha$ as function of $N_{conv}$ for $n_{std}=35$.}
\label{fig:35std}
\end{figure}

The truncation was done by erasing out the elements of a given set
with absolute value greater than $n_{std}\times \sigma$, where $\sigma$ is the standard deviation of the corresponding data.
It implies that, the smaller $n_{std}$, the fewer the elements remaining in the truncated set.
There is a range of $n_{std}$ when the simulation works.
For high $n_{std}$ (i.e., small truncation),
the length of the series is not large enough 
for observing the convergence to $\alpha=2$,
thought it seems to converge to a value significantly different from the value used for generating the set,
i.e., the case without convolution presented in fig.\ref{fig:sintruncar}.
An example is given in fig.\ref{fig:50std}.
On the other hand,
for low $n_{std}$ (i.e., large truncation),
the Kolmogorov-Smirnov test rejects that the data correspond to a 
stable distribution.
Neither it is a normal distribution,
but it converges very fast to $\alpha=2$.
An example is given now in fig.\ref{fig:10std}.
Therefore, reasonable truncation can be performed,
such that it can be obtained a series that for low $N_{conv}$
the test cannot reject them to be stable-distributed,
but with convolution converges to $\alpha=2$.
The corresponding example is given in fig.\ref{fig:35std}.

From these simulations it can be observed that 
the convergence
is not only slow, but also non uniform.
However, the larger the amount of elements in a given set,
the smaller the size of the irregularities.
If considering the whole population,
(a condition for the generalization of the Central Limit Theorem),
the convergence can be expected to be uniform.

\subsection{L\'evy to Gaussian crossover}

Since our sample is finite and this affect the estimation of the parameter $\alpha$, 
we further analyzed 
the transition from L\'evy to Gaussian regime by
following the procedure proposed in Ref.\cite{miranda}. 
We study the behavior of the excess kurtosis for convoluted samples, from $N_{conv}=0$ to $N_{conv}=3000$. 
The excess kurtosis 
\begin{equation}
k\equiv\frac{\langle(S_i-\mu)^4\rangle}{\langle(S_i-\mu)^2\rangle^2}-3 \, ,
\end{equation}
gives a statistical measure of the {\it heaviness} of the tail of a distribution with mean value $\mu$. 
A normal process shows zero excess kurtosis for the population, while it is positive for leptokurtic distributions like L\'evy-stable distributions. 

\begin{figure}[!ht]
	\centering
	\includegraphics[width=.7\linewidth]{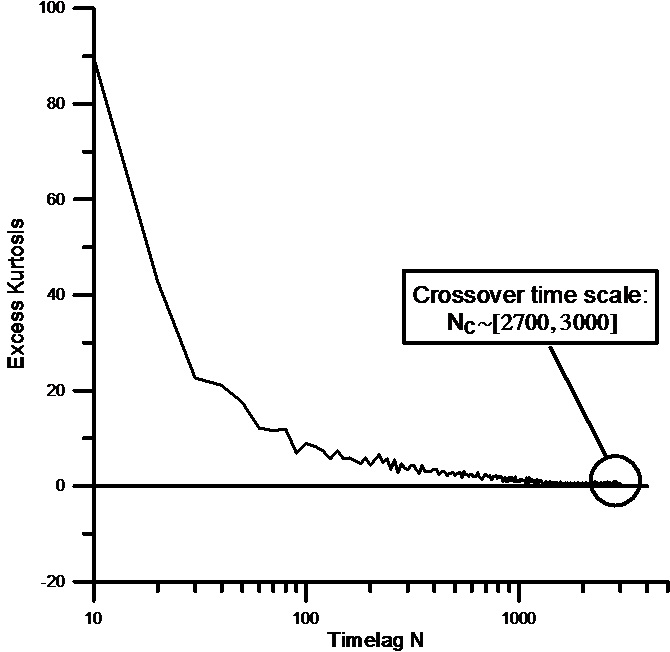}
	\caption{Excess kurtosis for the samples as a function of $N_{conv}$. The circle encloses the L\'evy to Gaussian crossover.}
	\label{fig:crossover}
\end{figure}

The results we obtained are presented in fig. \ref{fig:crossover}. 
It can be seen
that the transition between L\'evy and Gaussian regimes occurs for values of $N_{conv}$ between 2700 and 3000. 
This confirms the result discussed above for the fit of the convoluted data to a stable L\'evy distribution (see table \ref{tab:table1}), that gives a value of $\alpha=2$ for $N_{conv}=2700$. 
This way, the crossover time can be set equal to $N_c=2700$. 
Recalling that,
during the analyzed period (from January 1999 to December 2002), 
the average time between successive fluctuations is close to 20 seconds and that for the Mexican market one trading day is equal to 6.5 hours, we find that the Levy-Gaussian crossover is approximately 2.3 trading days.

\section{Serial dependence in the IPC data}
\label{sec:autocorr}

\begin{figure}[!ht]
	\centering
	\begin{subfigure}{.5\textwidth}
		\centering
		\includegraphics[width=\linewidth]{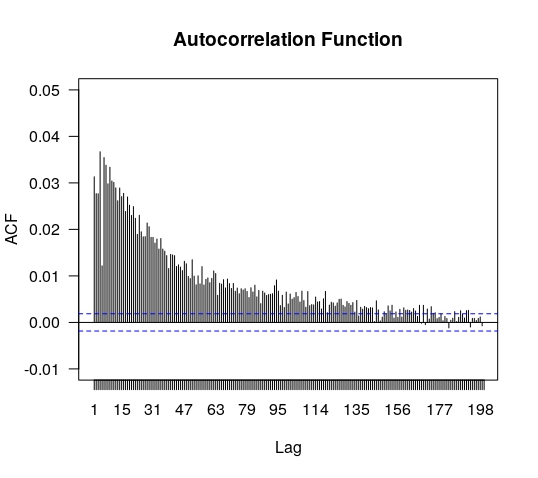}
	\end{subfigure}%
	\begin{subfigure}{.5\textwidth}
		\centering
		\includegraphics[width=\linewidth]{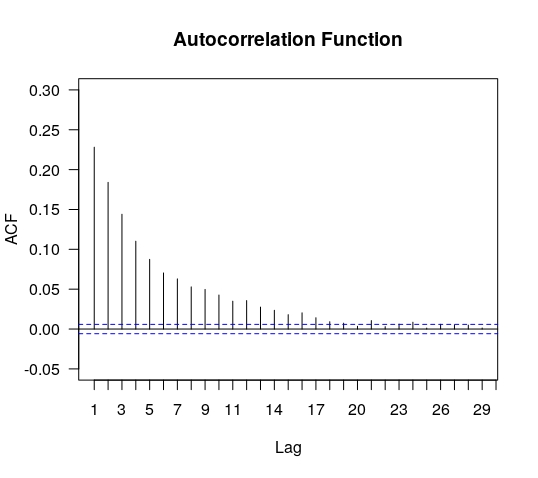}
	\end{subfigure}
		\caption{Autocorrelations for $N_{conv}=0$ (left) and $N_{conv}=10$ (right)}
		\label{fig:acfig1_2}
\end{figure}

\begin{figure}[!ht]
	\centering
	\begin{subfigure}{.5\textwidth}
		\centering
		\includegraphics[width=\linewidth]{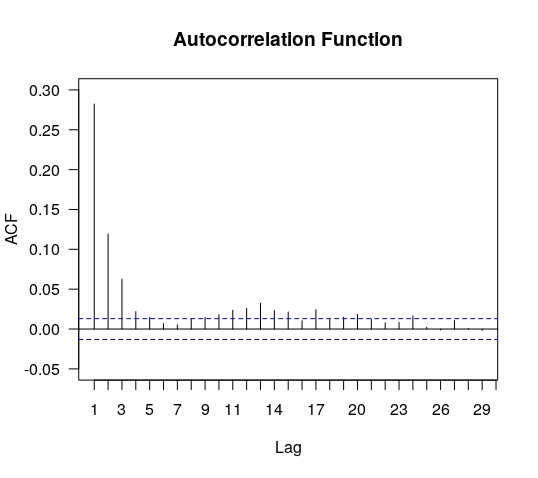}
	\end{subfigure}%
	\begin{subfigure}{.5\textwidth}
		\centering
		\includegraphics[width=\linewidth]{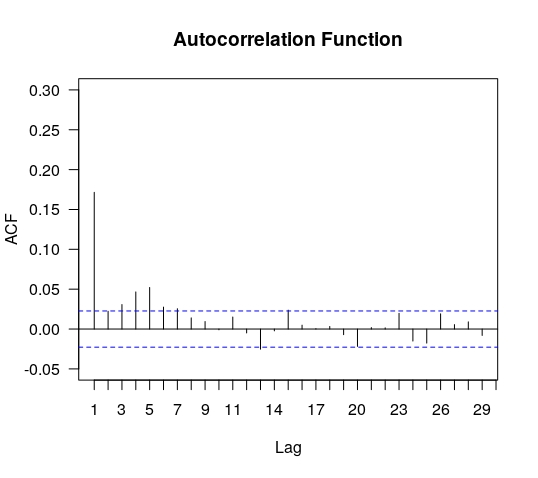}
	\end{subfigure}
\caption{Autocorrelations for $N_{conv}=50$ (left) and $N_{conv}=150$ (right)}
\label{fig:acfig3_5}
\end{figure}

From the previous section we concluded that for convolutions below $N_{conv}=70$ data does not fit neither a Gaussian nor any L\'evy distribution.
A common hypothesis for both cases is the data being
independent and identically distributed.
In figures \ref{fig:acfig1_2} and \ref{fig:acfig3_5} we present the
results for the analysis of autocorrelation for the set $\{S_k\}$.
In principle, the vertical axis would cover values from $-1$ (full anti-correlation) to $1$ (full correlation) and the values in the horizontal axis stand for the lag $\delta$, i.e., denote the correlation between $S_k$ and $S_{k+\delta}$.
The blue dashed lines in these figures
indicate approximate limits of correlation coefficients expected under a null hypothesis of uncorrelated data.
We successfully tested these limits using the simulations of truncated L\'evy distributions described in subsection \ref{ssec:convergence}.
As it can be observed the fluctuations exhibit positive autocorrelations, which are correspondingly diluted after convoluting the series. 
It suggests a mild serial dependence between fluctuations within an interval of about $48$ minutes.

\begin{figure}[!ht]
	\centering
	\begin{subfigure}{.5\textwidth}
		\centering
		\includegraphics[width=.9\linewidth]{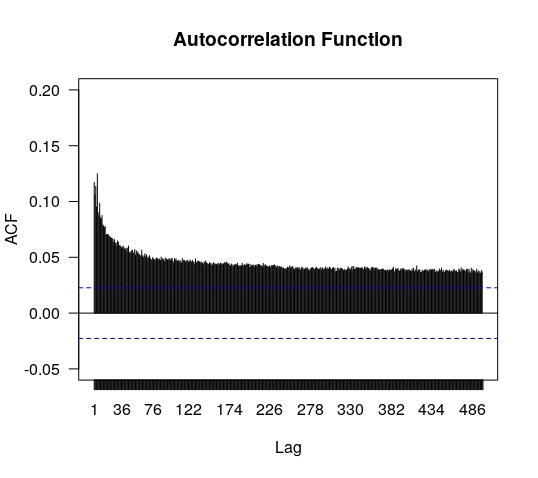}
	\end{subfigure}%
	\begin{subfigure}{.5\textwidth}
		\centering
		\includegraphics[width=.9\linewidth]{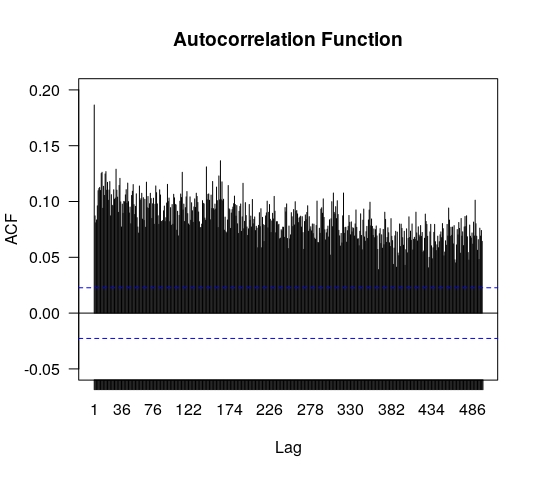}
	\end{subfigure}
	\caption{Autocorrelations for the absolute value of the log returns for $N_{conv}=0$ (left) and $N_{conv}=150$ (right)}
	\label{fig:volconv0_150}
\end{figure}

In figures  
\ref{fig:volconv0_150} 
are shown the autocorrelations for the series of absolute values of $S_k$ for $N_{conv}=0$ (left) and $N_{conv}=150$ (right).
This is a measure of volatility and
it exhibits long range serial dependence,
a fact consistent with findings reported for other markets
\cite{mantegna-stanley-book, cont}.

\begin{figure}[!ht]
	\centering
	\begin{subfigure}{.5\textwidth}
		\centering
		\includegraphics[width=.9\linewidth]{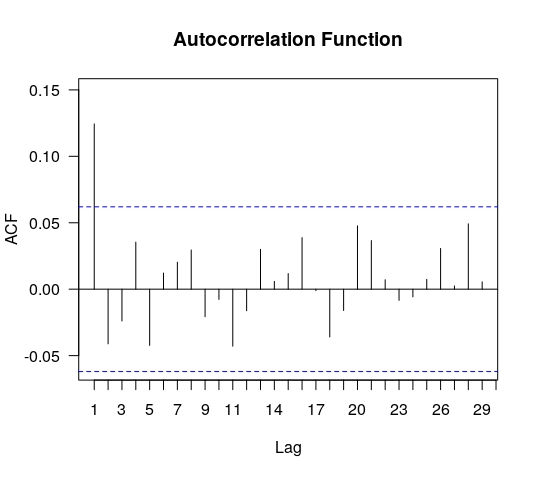}
	\end{subfigure}%
	\begin{subfigure}{.5\textwidth}
		\centering
		\includegraphics[width=.9\linewidth]{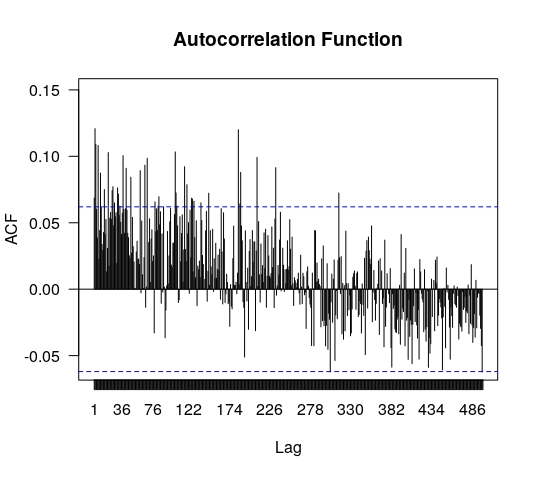}
	\end{subfigure}
	\caption{Autocorrelations for the log returns of the dayly closing (left) and the corresponding absolute value (right).}
	\label{fig:acfdaily}
\end{figure}

Finally, we present in figure
\ref{fig:acfdaily} 
the autocorrelations for the data and absolute value of the data of the daily closing set.
As it can be observed these last results are totally consistent with those obtained for the tick--to--tick set.

\section{Discussion}
\label{sec:discussion}

Our results suggest that the statistical description of the Mexican market strongly depends on the time scale of interest.
For the analyzed period, around 73\% of the IPC tick--to--tick data, 
sampled every $5.2$ seconds, 
shows no activity.
Fluctuations occur, in average, every $19.3$ seconds,
but the underlying probability is unlikely to follow 
a L\'evy-stable distribution, including the Gaussian.
Our analysis shows that one of the reasons could be that these data are not independent.
The fluctuations exhibit mild positive autocorrelations that persist for about $48$ minutes before falling below the level of noise.
This is, for instance, twice the value reported in Ref.\cite{mantegna-stanley-book} for the $S\&P$ $500$ index, sampled
at a 1 min time scale,
and also sometimes reported for various asset returns \cite{cont}. 
The autocorrelation in the IPC are diluted by convolution, i.e.,
the aggregation of elements of the set of fluctuations in blocks of length $N_{conv}$.
This is characteristic of random walks with short memory.
Conversely, in the case of a deterministic process with noise, 
even if autocorrelations are initially hidden,
they surface and get more noticeable with convolutions.
Therefore, this serial dependence seems to reflect less the internal mechanics of the market (due to the law of supply and demand) 
than the (complex, noisy) external action on it. 

It is worthy to note that sets obtained by convolution of the IPC tick--to--tick data
can be safely used to describe the activity at lower frequencies.
We tested this by comparing, for instance, the fit for $N_{conv}=1200$,
which corresponds to one trading day,
with the daily data from Yahoo finance for the same period (from January 1999 to December 2002).
From table \ref{tab:table1} we see that $\alpha=1.8693$,
while for the closure data we obtained $1.8646$.
Moreover, the analysis of serial dependence for the closure data (and their absolute value) presented in figures \ref{fig:acfdaily} is totally consistent with those obtained for the intra-day set described in the paragraph above.

Taking all of this into account,
the IPC fluctuations can be described
by L\'evy-stable distributions in the wide range of sampling interval
from 20 seconds up to two trading days.
After that time, data seems to obey normal probability distributions.
We would like to note that this value for the L\'evy to Gaussian crossover also differs from results obtained by other authors. For example, Mantegna and Stanley \cite{mantegna-stanley}, for the $S\&P 500$ (during the six year period from January 1984 to December 1989), estimated the crossover time to be of the order of one month; 
Matacz \cite{matacz}, for the Australian All Ordinaries share market index for the period 1993--1997,  found that the crossover time is approximately 19 trading days, and in Cuoto Miranda and Riera \cite{miranda}, a crossover of approximately 20 days was found for the Sao Paulo Stock Exchange Index in Brazil (IBOVESPA), during the 15 years period 1986–2000.

The differences outlined here with respect to the stylized facts found for other markets seems to indicate that in the given period the behavior of the Mexican market was atypical: the effects induced by external factors
(for instance, different kinds of expected announcements and unexpected
action of privileged agents) led the dynamics for periods of about an hour.
As this effect vanishes, 
significant (as compared to normal) fluctuations of the index from its average value were likely
within an interval of a couple of trading days,
but if analyzed over longer intervals,
data describes a usual random walk,
i.e., with the size of the fluctuations decaying exponentially.

Nevertheless, there are reasons for expecting the statistical description of the IPC to also depend on the period analyzed,
and the variation from period to period not just being given by the presence of critical events.
As mentioned before, in Ref.\cite{alfonso} we analyzed the daily closure data for the IPC covering the period from $04/09/2000$ to $04/09/2010$.  
For this whole period we obtained $\alpha=1.64$,
which is significantly lower than the value of $1.8646$ we obtained for
the daily data over the period $01/1999$ to $12/2002$ studied in this paper.
Both these values are still larger than,
for example, 
the one reported by 
Mantegna and Stanley \cite{mantegna-stanley} for the $S\&P 500$ about a decade earlier.

\section{Acknowledgments}

This research was supported by the Sistema Nacional de Investigadores (M\'exico).
It was also partially funded by a grant from the Consejo Nacional de Ciencia y Tecnología of Mexico (SEP--CONACyT) CB-284482.
D. G-R. thanks the support of the Facultad de Ciencias, Universidad de Colima, where most of her contribution was done.


\begin{thebibliography}{99}

\bibitem{fama}
E. F. Fama, J. Finance, {\bf 46} 1575–613 (1991)

\bibitem{mantegna-stanley-book}
R.~N.~Mantegna and H.~E.~Stanley, 
{\em An introduction to econophysics: correlations and complexity in finance.}, 
(Cambridge University Press, New York, (1999))

\bibitem{levy}
P.~L\'evy, {\em Calcul des Probabilites}, (Paris : Gauthier-Villars et Cie, (1925))

\bibitem{mandelbrot}
B.~B.~Mandelbrot, Journal of Business {\bf 36}, 394 (1963).

\bibitem{mantegna-stanley}
R.~N.~Mantegna and H.~E.~Stanley, Nature {\bf 376} 46 (1995).

\bibitem{gopikrishnan} P. Gopikrishnan, V. Plerou, L.~A.~N.~Amaral, M. Meyer, and H. E. Stanley, 
Phys. Rev. E 60, 5305–5316 (1999) 

\bibitem{cont} Cont, R., Quantitative Finance, {\bf 1}: 223-236, (2001)

\bibitem{alfonso}
L. Alfonso, R. Mansilla, and C.A. Terrero-Escalante,  Physica A: Statistical Mechanics and its
Applications {\bf 391}, 2990–2996 (2012).

\bibitem{mantegna-stanley-PRL}
R.~N.~Mantegna and H.~E.~Stanley, Physical Review Letters, {\bf 73} 22, 2946 (1994).

\bibitem{koponen} I. Koponen, Phys. Rev. E {\bf 52}, 1197 (1995)

\bibitem{andersen}
T. G. Andersen, Journal of Business and Economic Statistics 18, no. 2 (2000): 146-53. doi:10.2307/1392552.

\bibitem{weron}
R.~Weron, International Journal of Modern Physics C {\bf 12} 2 (2001).

\bibitem{taqqu}
G.~Samorodnitsky and M.~S.~Taqqu, {\em Stable Non-Gaussian Random Processes. Stochastic Models with Infinite Variance.}, 
(Chapman and Hall, New York - London (1994))

\bibitem{cont-potters-bouchaud} 
R. Cont, M. Potters, J. P. Bouchaud, 
{\em Scale Invariance and Beyond}. 
(Centre de Physique des Houches, vol 7. Springer, Berlin, Heidelberg (1997)) 

\bibitem{matacz}
A. Matacz, International Journal of Theoretical and Applied Finance, {\bf 3} (01), 143-160 (2000).

\bibitem{nolan}
http://fs2.american.edu/jpnolan/www/stable/stable.html

\bibitem{miranda}
L. C. Miranda, and R. Riera,  Physica A: Statistical Mechanics and its Applications, {\bf 297} (3-4), 509-520 (2001).

\end{thebibliography}
\end{document}